\documentclass[reprint, aps,superscriptaddress,amsmath,amssym,prl]{revtex4-2}

\usepackage{bm}
\usepackage{float}
\usepackage{graphicx}
\usepackage[usenames,dvipsnames]{color}
\usepackage[normalem]{ulem}
\usepackage[svgnames]{xcolor}
\usepackage{bm}
\usepackage{multirow}
\usepackage{float}
\usepackage{titlesec}
\usepackage[utf8]{inputenc}
\usepackage[colorlinks,linkcolor=blue,citecolor=blue,urlcolor=blue]{hyperref}
\begin{document}
	\title{Angle-dependent planar thermal Hall effect by quasi-ballistic phonons in black phosphorus}
	
	\author{Xiaokang Li}
	\email{lixiaokang@hust.edu.cn}
	\affiliation{Wuhan National High Magnetic Field Center, School of Physics, Huazhong University of Science and Technology,  Wuhan  430074, China}
	
	\author{Xiaodong Guo}
	\affiliation{Wuhan National High Magnetic Field Center, School of Physics, Huazhong University of Science and Technology,  Wuhan  430074, China}
	\affiliation{Laboratoire de Physique et d'Etude de Mat\'{e}riaux (CNRS)\\ ESPCI Paris, PSL Research University, 75005 Paris, France }
	
	\author{ Zengwei Zhu}
	\email{zengwei.zhu@hust.edu.cn}
	\affiliation{Wuhan National High Magnetic Field Center, School of Physics, Huazhong University of Science and Technology,  Wuhan  430074, China}
	\author{Kamran Behnia}
	\email{kamran.behnia@espci.fr}
	\affiliation{Laboratoire de Physique et d'Etude de Mat\'{e}riaux (CNRS)\\ ESPCI Paris, PSL Research University, 75005 Paris, France }
	
	\begin{abstract}
		The origin of the phonon thermal Hall effect in insulators is a matter of ongoing debate. The large amplitude of the signal in an elemental non-magnetic solid, such as black phosphorus (BP), calls for a minimal mechanism not invoking the spin degree of freedom. Here, we show that a longitudinal heat flow generates a transverse temperature gradient in BP even when the magnetic field, the heat current and the thermal gradient lie in the same plane. The phonon mean-free-path is close to the sample thickness. Therefore, it is unlikely  that scattering by point-like symmetry-breaking defects play a major role. We show that the  angular dependence of the signal can be mapped to the sum of two sinusoidal components each peaking when the magnetic field is parallel to a high symmetry. We propose that anharmonicity may play a major role and argue that the magnetic field can exert a torque on electric dipolar waves traveling with phonons.
		
	\end{abstract}
	\maketitle

	\section{1. Introduction}
	The thermal conductivity is a second rank tensor linking  the heat current density and the temperature gradient vectors. The thermal Hall effect (THE) refers to non-zero off-diagonal components of this tensor, $\overline{\kappa}$. Its  origin in insulators has attracted much recent experimental \cite{Strohm2005,Onose2010,Chen2019LVO,Kasahara2018,Grissonnanche2019,Li2020,Grissonnanche2020,Boulanger2020,Akazawa2020,Sim2021,Chen2022,Uehara2022,Jiang2022,Lefran2022,Li2023,Sharma2024,Meng2024} and theoretical \cite{Sheng2006, Kagan2008,Zhang2010, Qin2012,Agarwalla2011,Chen2020,Flebus2022,Guo2022,Mangeolle2022,Chen2024review} interest.
	
	The planar thermal Hall effect refers to a configuration in which the three relevant vectors (heat current, temperature gradient and magnetic field) lie in the same plane. This was first observed in the Kitaev spin liquid candidate $\alpha$-RuCl$_3$ \cite{Kasahara2018, Yokoi2021, Czajka2021, Czajka2023}. More recently, it was also observed in other solids, where the thermal Hall signal is attributed to phonons \cite{Chen2024,barthélemy2023planar,Jin2024}. The persistence of the signal, when it is  forbidden by crystal symmetry, was attributed to defects breaking the local symmetry  \cite{barthélemy2023planar}.

	\begin{figure*}[ht]
		\includegraphics[width=1.0\linewidth]{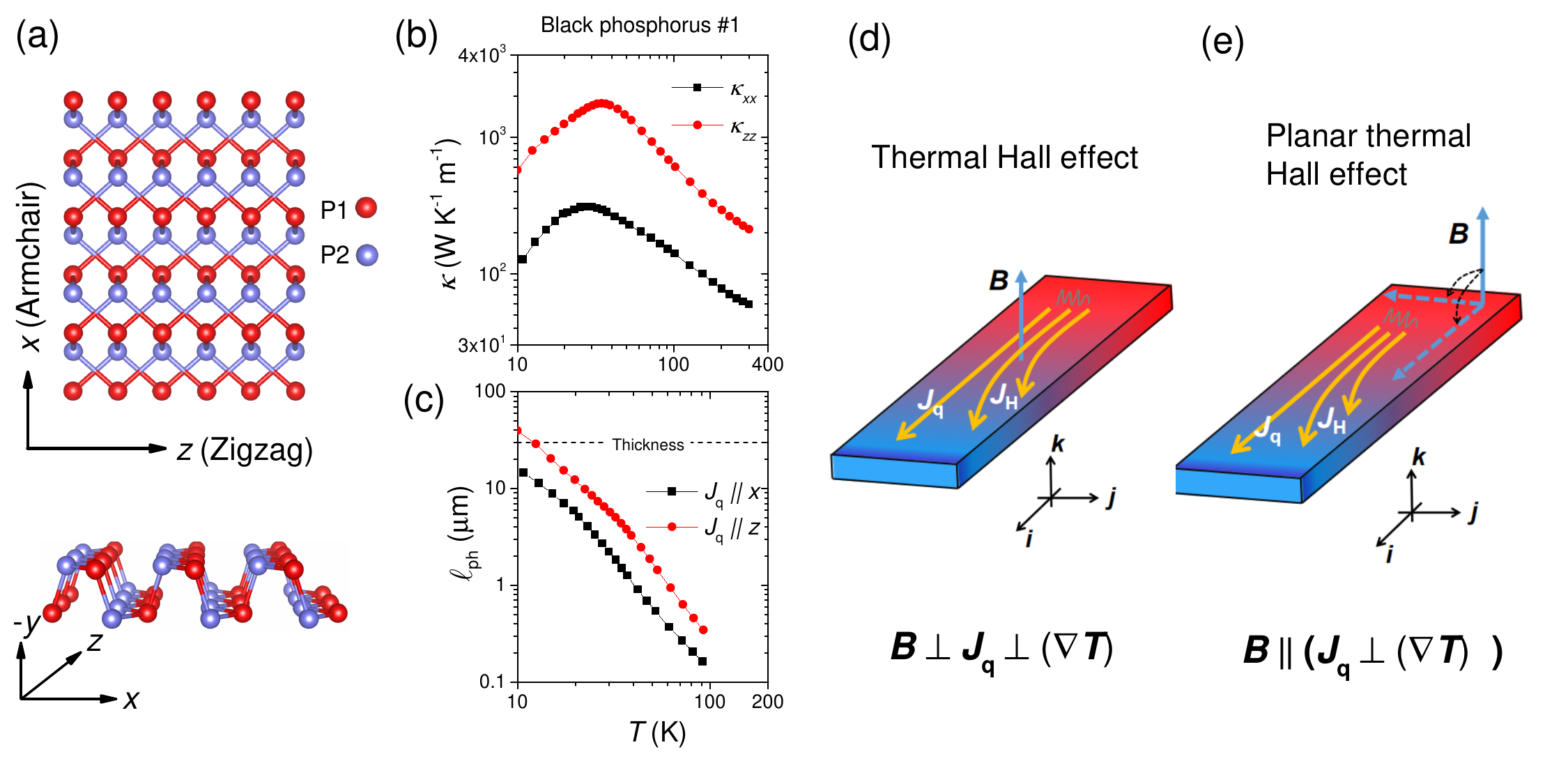}
		\caption{Crystal structure, phonon mean free path and two versions of the thermal Hall effect.(a) Top (two layers) and side (single layer) view of the lattice of black phosphorus (BP). P atoms, marked in blue and red, belong to atomic sites with distinct environments. 
			(b) Longitudinal thermal conductivity (from previous work~\cite{Li2023}) along the armchair ($\kappa_{xx}$) and the zigzag ($\kappa_{zz}$) orientation, at least four times of anisotropy can be seen. (c) The temperature dependence of mean free path of phonon ($\ell_{\mathrm{ph}}$) in BP for two in-plane orientations ($J_\mathrm{q}||z$ axis and $J_\mathrm{q}||x$ axis ), extracted from the longitudinal thermal conductivity, the sound velocity and the specific heat. It shows an increasing behavior before 10 K with cooling and is comparable to the sample thickness around 20 K, the peak temperature of $\kappa_{ii}$.  (d) The schematic for measuring the thermal Hall effect, with the magnetic field ($B$) perpendicular to the longitudinal heat current ($J_\mathrm{q}$) and the transverse thermal gradient ($(\nabla T)_{\perp}$). The thermal Hall conductivity tensor is denoted as $\kappa_{ij}^k$. (e) The schematic for measuring planar thermal Hall effect, with the magnetic field ($B$) rotating from the orthogonal to the parallel direction of the $J_\mathrm{q}$ or the $(\nabla T)_{\perp}$. Two types of planar thermal Hall conductivity tensors are denoted as $\kappa_{ij}^i$ and $\kappa_{ij}^j$ respectively.  }
		\label{fig:config}
	\end{figure*}
	
	\begin{figure*}[ht]
		\includegraphics[width=1.0\linewidth]{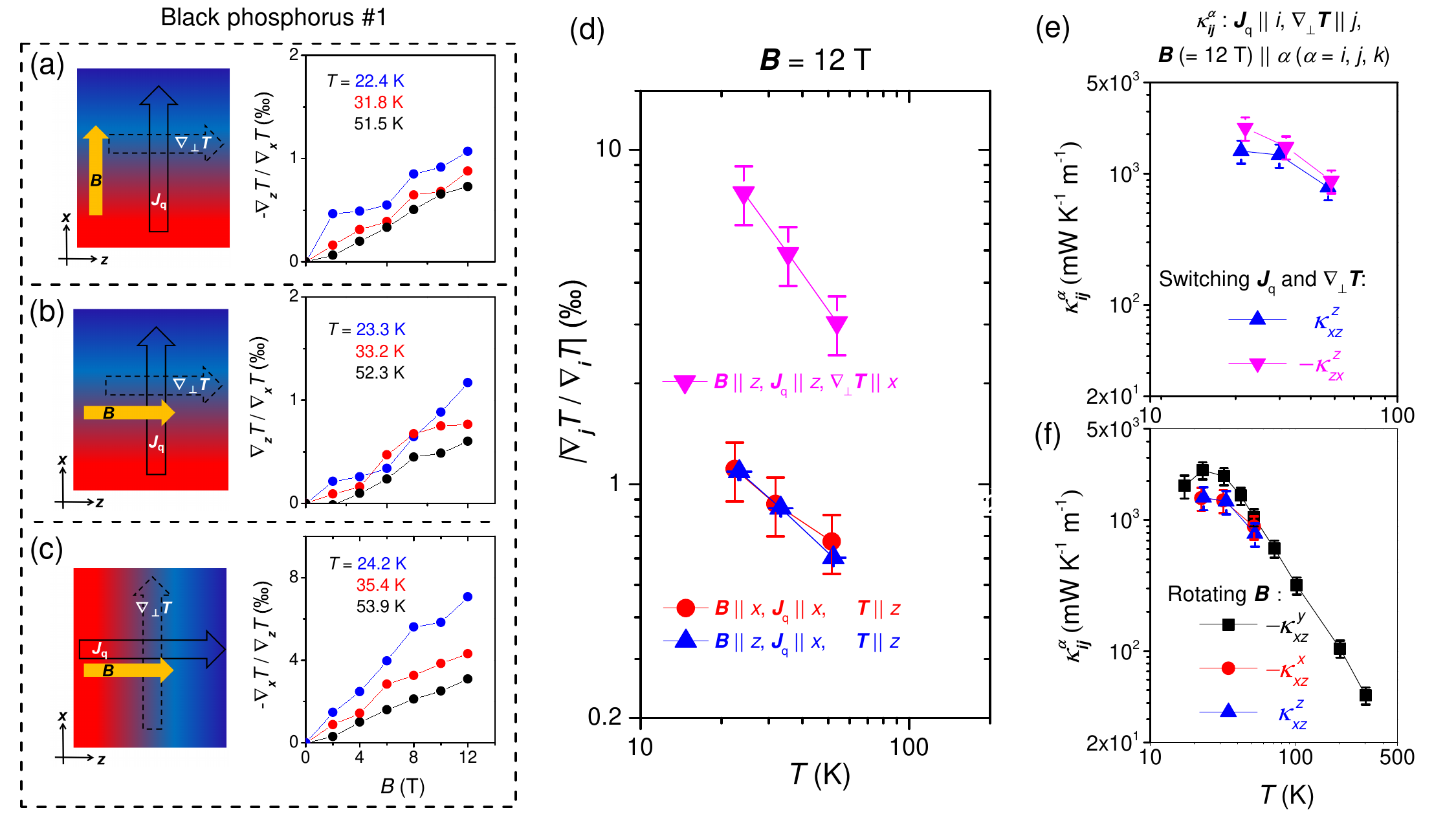}
		\caption{Planar thermal Hall data. (a) The field dependence of thermal Hall angle $\nabla_j T / \nabla_i T$ , with $J_\mathrm{q}$ and $B$ along the $x$ axis, $(\nabla T)_{\perp}$ along the $z$ axis, at three typical temperatures. This allows to quantify $\kappa_{xz}^x$. (b)  Same for $J_\mathrm{q}$ along the $x$ axis, $B$ and $(\nabla T)_{\perp}$ along the $z$ axis.  This allows to quantify $\kappa_{xz}^z$. (c)  Same for $J_\mathrm{q}$ and  $B$  along the $z$ axis and $(\nabla T)_{\perp}$ along the $x$ axis.  This allows to quantify $\kappa_{zx}^z$. (d) The temperature dependence of the thermal Hall angle for three different configurations at 12 T. All measurements are performed on the same sample. Note that the thermal Hall angle, $\nabla_j T /\nabla_i T$ is almost unchanged when the orientations of $B$ rotates from $J_\mathrm{q}$ to $(\nabla T)_{\perp}$. On the other hand, it changes by a factor of six times when the orientations of $J_\mathrm{q}$ and $(\nabla T)_{\perp}$ permute with each other. This difference combined with the anisotropy in longitudinal thermal conductivity warrants Onsager reciprocity: $\kappa_{xz}^z=- \kappa_{zx}^z$, as seen in (e). (f) Comparison of thermal Hall conductivity $\kappa_{xz}$ with the magnetic field $B$ rotating from the orthogonal ($y$ axis, from previous work~\cite{Li2023} with the same sample) to the parallel direction of the $J_\mathrm{q}$ ($x$ axis) or the $(\nabla T)_{\perp}$ ($z$ axis).}
		\label{fig:raw}
	\end{figure*}
	
	\begin{figure*}[ht]
		\includegraphics[width=0.9\linewidth]{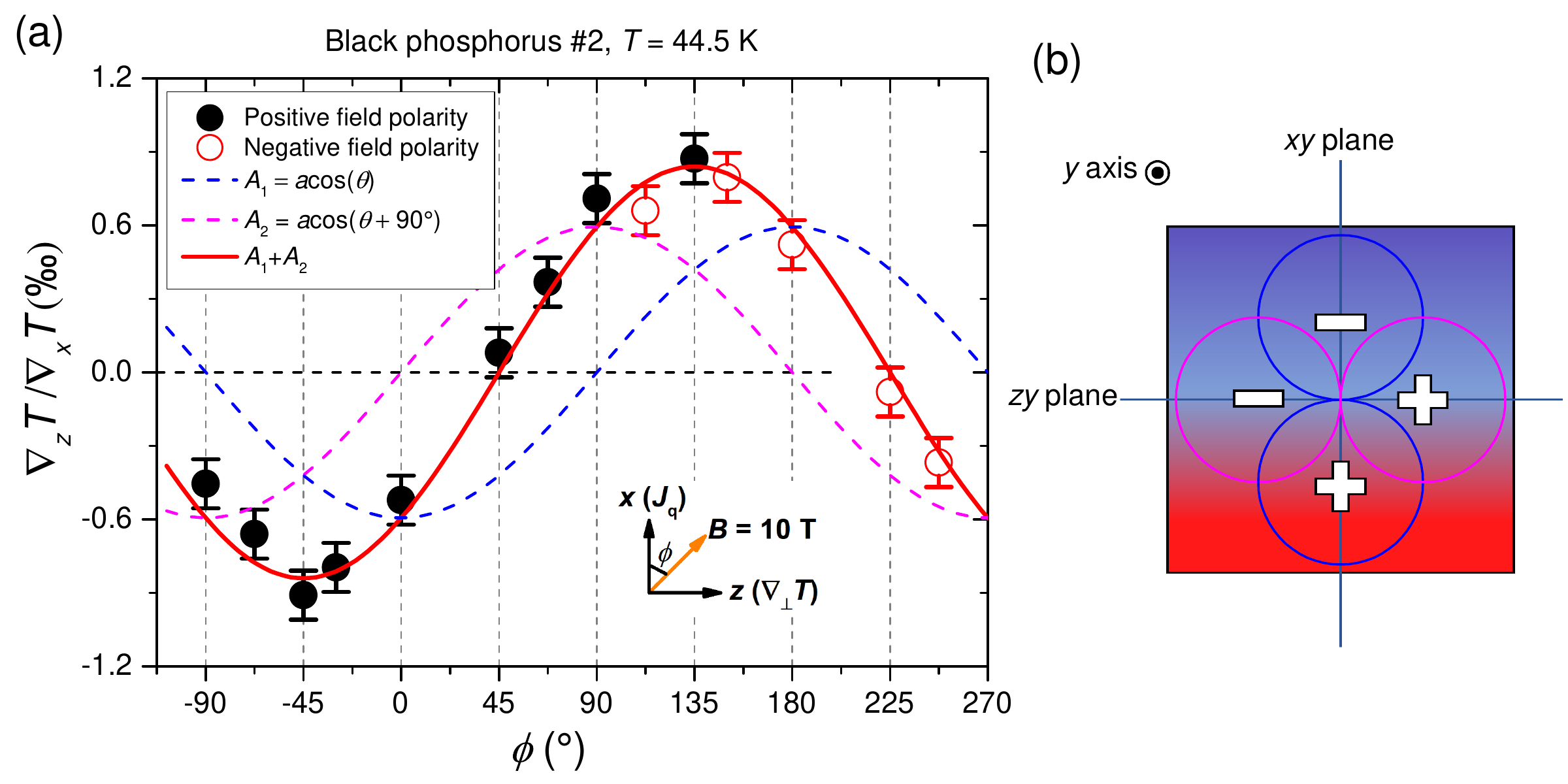}
		\caption{Angle dependence of the thermal Hall angle in the planar configuration. (a) Angular dependence of $\nabla T_z/\nabla T_x$ at a constant temperature and constant magnetic field. The red line is a cosine fit to the data, and the sum of two blue lines representing cos$\phi$ and cos$(\phi+\pi/2)$. Each of these components peaks along one of the two high-symmetry axis and vanishes along the other symmetry axis. Suggesting that the planar thermal Hall signal consists of two distinct contributions each along one high-symmetry axis. (b) A schematic representation of  the two high-symmetry axes and the polarities implied by the observation of an odd signal when the field is parallel to each of them. Colors represent the temperature profile.}
		\label{fig:angular}
	\end{figure*}
	
	\begin{figure}[ht]
		\includegraphics[width=1\linewidth]{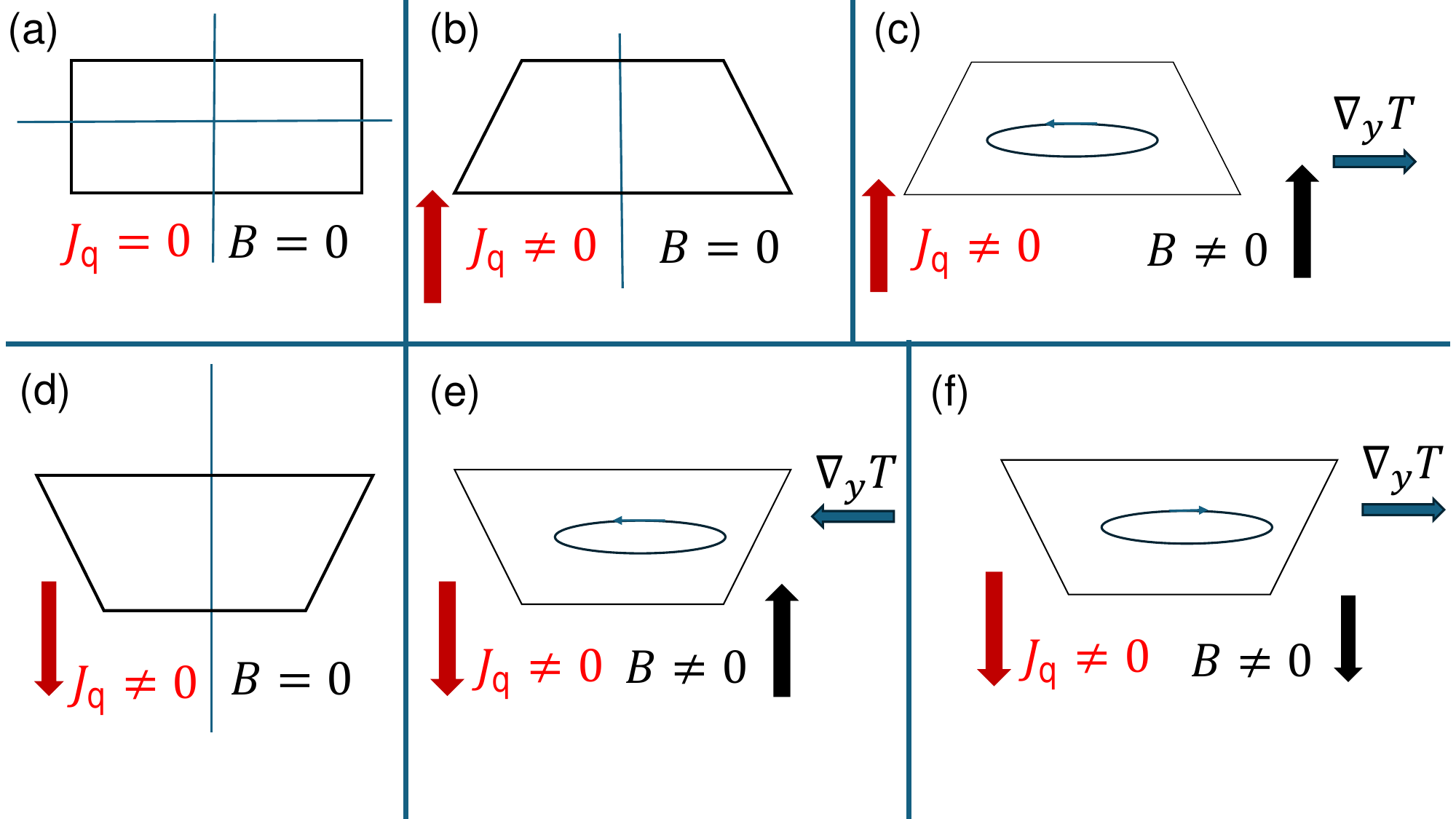}
		\caption{Unit cell in presence of a heat current and magnetic field. (a) The pristine unit cell without the heat current and the magnetic field. (b) The unit cell in presence of a heat current $J_\mathrm{q}$, It changed from a rectangle to a trapezoid. The only mirror plane which survives is the one parallel to the heat current. Which vanishes too in presence of a magnetic field $B$, as seen in (c). The mirror planes not withstanding, a planar Hall response can emerge. (d) Changing the orientation of the heat current will change the way the primitive cell is distorted. In this case, applying a magnetic field along the same orientation will generate a transverse thermal gradient of opposite sign (e). On the other hand, inverting the orientation of the magnetic field will double flip the orientation of the transverse thermal gradient (f).
		}
		\label{fig:unit cell}
	\end{figure}
	
	\begin{figure}[ht]
		\includegraphics[width=1\linewidth]{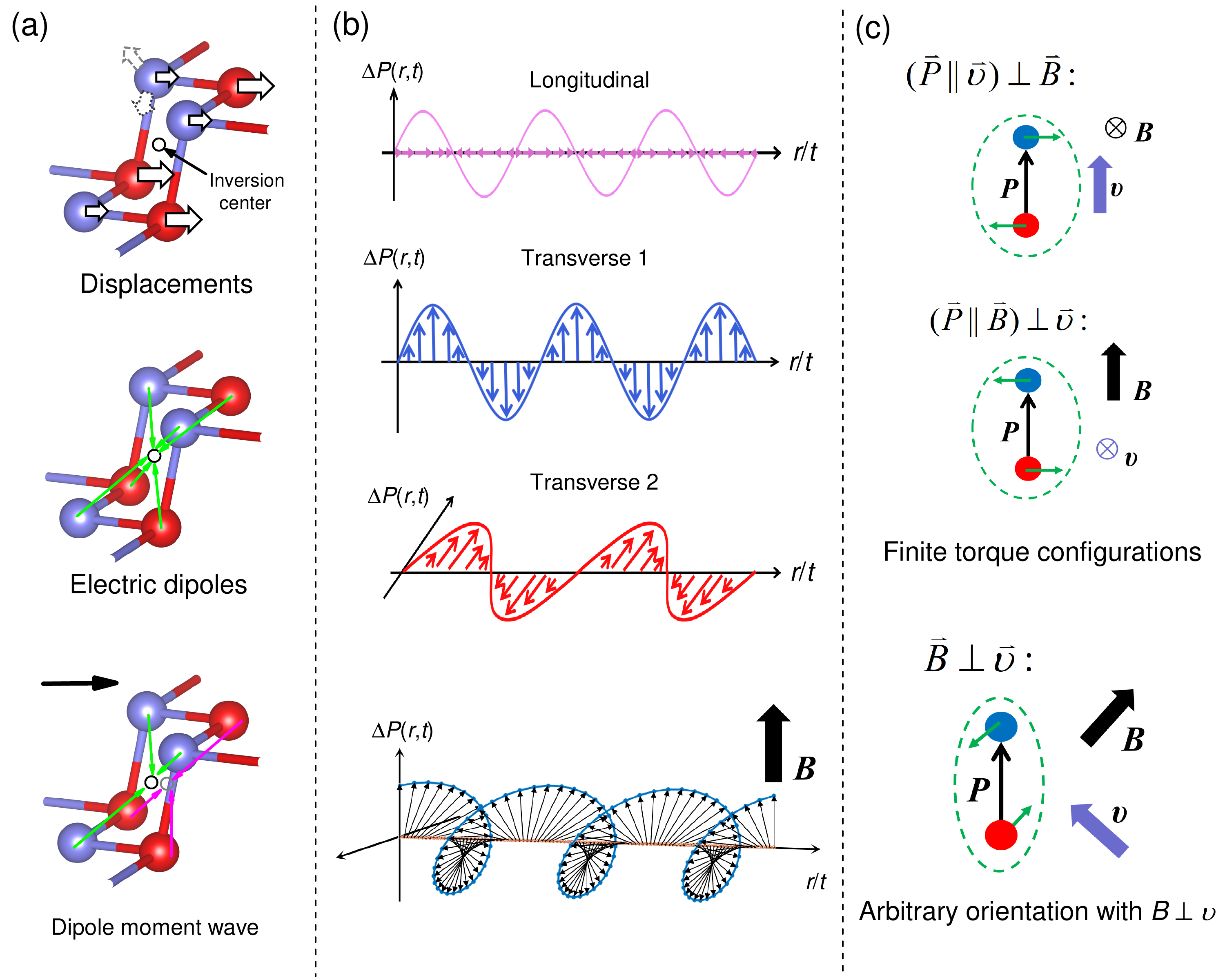}
		\caption{From atomic displacements to dipolar moments. 
			(a) Phosphorus atoms surrounding an inversion center. Their displacements associated with an acoustic mode (top) will affect the local electric dipole moments (middle), generating a net dipole moment (bottom).  (b)  Phonon modes are associated with waves of dipolar moments. A static magnetic field will exert a torque on the dipole moment traveling with the sound wave.  (c)  This torque ($\tau= \overrightarrow{P}\times (\overrightarrow{B}\times\overrightarrow{v})$) will survive only when the magnetic field is oriented perpendicular to the wave velocity. Green arrows show the orientation of the Lorenz forces on the two poles of the dipole.
		}
		\label{fig:dipoles}
	\end{figure}

	Here, we report on the observation of a planar Hall effect in black phosphorus(P), an elemental insulator with ballistic phonons \cite{Machida2018}. The thermal conductivity tensor has unequal diagonal components ($\kappa_{zz} \simeq 5 \kappa_{xx}$). Nevertheless, the off-diagonal components match each other ($\kappa_{xz}(B)\simeq\kappa_{zx}(-B)$), as expected by Onsager reciprocity. We quantify the variation of the thermal Hall signal as a function of the in-plane orientation of the magnetic field and find that it displays a twofold oscillation with minimum and maximum along one of the two diagonals of the $xz$ plane of the puckered honeycomb lattice. This indicates that the signal is the sum of two sinusoidal contributions along two high-symmetry axes. A quantitative account of our observation is missing. Nevertheless, we argue that because of the finite thermal expansion, a temperature gradient allows the emergence of a signal forbidden in thermodynamic equilibrium. Electric dipole moments travelling with sound can couple to magnetic field and generate a thermal Hall signal in planar configuration.
	
	\section{2. Materials and methods}
	Detailed materials and methods can be found in the supplementary materials, including sample details, thermal transport measurement methods.
	
	\section{3. Results}
	Fig.~\ref{fig:config}a shows the crystal structure of black P. Identical phosphorus atoms located on two distinct sites are marked in blue and red. Each layer is a puckered honeycomb lattice in the $xz$ crystallographic plane, where the $x$ and $z$ axes correspond to the armchair and zigzag orientations. As seen in Fig.~\ref{fig:config}b,  the longitudinal thermal conductivity, as found previously \cite{Machida2018,Sun2017}, is significantly different along the two orientations.  Along the zigzag orientation, thermal conductivity is much larger than along the armchair orientation in the low temperature limit. The sound velocity shows a similar but attenuated anisotropy (9.6 km/s along $z$-axis \textit{vs.} 4.6 km/s along the $x$-axis) \cite{Fujii1982}. As a consequence, the phonon mean free path  $\ell_{\mathrm{ph}}(T)$, shown in Fig.~\ref{fig:config}c, is also anisotropic. It is twice longer along the $z$-axis and approaches the sample thickness (30 $\mu$m) around 20 K. Thus, at this temperature phonons become quasi-ballistic and accordingly thermal conductivity  becomes size dependent  \cite{Machida2018}. 
	
	A Hall response refers to a signal odd in magnetic field and oriented perpendicular to the applied current. Designating the transverse temperature gradient by $(\nabla T)_{\perp}$, the applied longitudinal heat current  by $J_\mathrm{q}$ and the magnetic field by $B$, the standard thermal Hall configuration corresponds to $B \perp J_\mathrm{q} \perp (\nabla T)_{\perp}$. It is sketched in panel d and was the subject of our previous study \cite{Li2023}. Panel e shows the configuration for planar thermal Hall effect, which corresponds to $B ~||~ (J_\mathrm{q} \perp (\nabla T)_{\perp})$. We use $\kappa_{ij}^k$ to designate the  $\overline{\kappa}$  component corresponding to $J_\mathrm{q}$ along the $i$-axis, $(\nabla T)_{\perp}$ along the $j$-axis and  the magnetic field along the $k$-axis.
	
	Fig.~\ref{fig:raw}a--c shows the planar thermal Hall data for three different configurations. In each panel, the orientations of the three relevant vectors ($B$, $J_\mathrm{q}$ and $(\nabla T)_{\perp}$) are sketched and the field dependence of the ratio of the transverse to longitudinal temperature gradients ($\nabla_j T /\nabla_i T$) is shown.
	
	Fig.~\ref{fig:raw}d compares the temperature dependence of the thermal Hall angle in three planar configurations at 12 T. It is finite in the three planar configurations, but there is a large difference in amplitude when one permutes the orientations of  $J_\mathrm{q}$ and $(\nabla T)_{\perp}$. Since the longitudinal thermal conductivity is anisotropic by the same factor, this difference warrants an equality between the absolute values of  $\kappa_{xz}^z$ and  $\kappa_{zx}^z$, as expected by Onsager reciprocity.

	The thermal Hall angle (Fig.~\ref{fig:raw}d) combined with the longitudinal thermal conductivity (Fig.~\ref{fig:config}b) leads to the planar thermal Hall conductivity, which is shown in Fig.~\ref{fig:raw}e and f. The thermal Hall response is finite in four distinct configurations. 
	We can see not only the validity of the Onsager reciprocity (Fig.~\ref{fig:raw}e), but also the very small influence of the orientation of magnetic field on the amplitude of $\kappa_{xz}$ (Fig.~\ref{fig:raw}f).

	We then proceeded to quantify the angle dependence of this planar Hall signal at fixed temperature as the magnetic field rotated in the $xz$ plane. The raw data, shown in the Supplementary material, shows a field-linear response for all explored angles. The measurements were performed at 44.5 K.This temperature was chosen to strike a balance between proximity to the peak temperature and the optimal sensitivity of the thermocouples. As seen in Fig.~\ref{fig:angular}a, the signal exhibits a twofold oscillation. It peaks (with opposite signs) at $\phi=-\pi/4$ and $\phi=3 \pi/4$, i.e., along one of the diagonals of the $xz$ plane, and vanishes along the other diagonal.

    This intriguing feature would find a straightforward interpretation if the signal is the sum of two contributions of equal amplitude shifted by $\pi/2$, one following cos$\phi$ (peaking along the $x$-axis and vanishing along the $z$-axis) and another following cos$(\phi+\pi/2)$ (peaking along the $z$-axis and vanishing along the $x$-axis), indicated by dashed lines in Fig.~\ref{fig:angular}a. As sketched in Fig.~\ref{fig:angular}b, each odd-field contribution would be bounded to one mirror plane.
	
	\section{4. Discussion and conclusion}
	Thus, we detect in a crystal belonging with the $mmm$ point group symmetry, there is a planar thermal Hall signal with bi-axial symmetry. Let us note that at peak temperature of this thermal Hall signal,  the longitudinal thermal conductivity of black P is as large as $\kappa_{\mathrm{p}}\simeq 2000$ WK$^{-1}$m$^{-1}$. This is to be compared  with cases such as $\alpha$-RuCl$_3$ ($\kappa_{\mathrm{p}}\simeq$ 3 WK$^{-1}$m$^{-1}$) \cite{Lefran2022}, Na$_2$Co$_2$TeO$_6$ ($\kappa_{\mathrm{p}}\simeq$ 10  WK$^{-1}$m$^{-1}$) \cite{Chen2024} or Y-kapellasite ($\kappa_{\mathrm{p}}\simeq$ 2  WK$^{-1}$m$^{-1}$) \cite{barthélemy2023planar}. The thermal conductivity of black phosphorus is three orders of magnitude larger, with a phonon mean free path close to the sample thickness. These features strongly suggest that the thermal transport mechanism in black phosphorus is dominated by boundary scattering and phonon-phonon scattering, rather than being influenced by local defect scattering. However, we cannot totally exclude a minor role played by local defects.
	
	In anisotropic dense medium, thermal diffusivity is a tensor, $\overline{D}$. The heat equation becomes:
	\begin{equation}
		\frac{\partial T}{\partial t}= \overrightarrow{\nabla}\cdot (\overline{D} \overrightarrow{\nabla} T)
		\label{Eq_heat}
	\end{equation}
	
	Consider now the energy continuity equation, which relates the energy flux to specific heat per volume, $C$:
	\begin{equation}
		\overrightarrow{\nabla}\cdot \overrightarrow{J_\mathrm{q}}+C\frac{\partial T}{\partial t} =0
		\label{Eq_Cont}
	\end{equation}
	
	The combination of the two previous equations leads to:
	
	\begin{equation}
		\overrightarrow{J_\mathrm{q}}=  - C \overline{D} \overrightarrow{\nabla} T
		\label{Eq_Fourrier}
	\end{equation}
	
	The thermal conductivity tensor $\overline{\kappa}$, is thus the thermal diffusivity tensor multiplied by a scalar: $\overline{\kappa}= \overline{D} C$. Therefore, off-diagonal components in $\overline{\kappa}$ are proportional to their counterparts in $\overline{D}$. The latter is the product of carrier velocity and carrier mean free path. Sound velocity, $v_\mathrm{s}$, and phonon mean free path $\ell_{\mathrm{ph}}$ (see Fig.~\ref{fig:config}c) are both anisotropic at zero magnetic field. The issue is to find a way for magnetic field to skew one or both of these tensors off the symmetry axes.
	
	Recent theoretical studies have scrutinized intrinsic phonon band \cite{Yang2020} incorporated with energy magnetization contribution \cite{Qin2011,Qin2012} and phonon angular momentum acquired through interaction with the spins \cite{Zhang2014,Lifa2015} or with orbital motion of ions \cite{Juraschek2019}.
	
	Absent magnetism,  the interplay between electric polarization and phonons deserves scrutiny. The computed map of charge distribution is strongly orientational \cite{Hu2016}. Dipole-active phonon modes has been detected by infrared spectroscopy \cite{SUGAI1985753}. These are optical phonons and not heat-carrying acoustic modes which interest us. The lowest optical phonon frequency of black phosphorus is around 140 cm$^{-1}$ \cite{SUGAI1985753} (see the Supplementary material for more details), much larger than the acoustic phonons. However, for the high temperature signals, the indirect contribution of optical phonons through coupling with acoustic phonons may not be negligible. Nevertheless, any phonon breaking the inversion symmetry will generate an electric dipolar wave. This feature highlighted in the context of ferrons, the elementary excitation of a ferroelectric solid \cite{Brandi2023,Bauer2023}, persists even in a paraelectric insulator.
	
	Black P has an anisotropic thermal expansion with opposite signs along armchair and zigzag crystalline orientations \cite{SUN20162098}. In presence of a temperature gradient,  each unit cell is slightly distorted (See Fig.~\ref{fig:unit cell}a, b and d). This feature, driven by unavoidable anharmonicity in any solid \cite{Ashcroft76}, is the fundamental reason that a signal forbidden in thermodynamic equilibrium is allowed in presence of a temperature gradient. It also paves the way for chirality in presence of the magnetic field (See Fig.~\ref{fig:unit cell}c, e and f).
	
	Let us now draw a tentative picture of how the magnetic field may couple to heat-carrying phonons. The inversion center of the pristine unit cell is not an atomic site and atomic displacements associated  with phonons generate electric dipole moments (Fig.~\ref{fig:dipoles}a). Therefore, acoustic phonons breaking inversion centers are traveling electric dipolar moments capable of coupling to a static magnetic field  (Fig.~\ref{fig:dipoles}b). Electric dipole wave and its evolution under magnetic field can be detected via methods such as $P$-$E$ response and second harmonic generation directly in the future. The torque exerted by a static magnetic field, $\overrightarrow{B}$, on an electric dipole of $\overrightarrow{P}$ moving with a velocity of $\overrightarrow{v}$ is : $\overrightarrow{\tau}= \overrightarrow{P}\times (\overrightarrow{B}\times\overrightarrow{v})$ \cite{Mungan2008}. Let us rewrite it as:
	\begin{equation}
		\overrightarrow{\tau}= (\overrightarrow{P}\cdot\overrightarrow{v}) \overrightarrow{B}- (\overrightarrow{P}\cdot\overrightarrow{B}) \overrightarrow{v}
		\label{dipole}
	\end{equation}
	
	Thus, the torque due to the Lorenz force exerted  on each of the two poles is finite when the magnetic field is oriented perpendicular to the orientation of the dipole propagation  (Fig.~\ref{fig:dipoles}c).  These are ingredients for a scenario for planar thermal Hall signal in a non-magnetic insulator (see the Supplementary material for a discussion of the expected angular variation).  A rigorous treatment is beyond the scope of this paper, but torque is known to be responsible for changes in angular momentum, so a quantitative theoretical analysis could involve either the phonon angular momentum \cite{Zhang2014,Yao2025} or the phonon Berry curvature \cite{Zhang2010}. However, let us compare the order of magnitude of the expected and the measured signals.
	
\begin{figure}[ht]
		\includegraphics[width=1\linewidth]{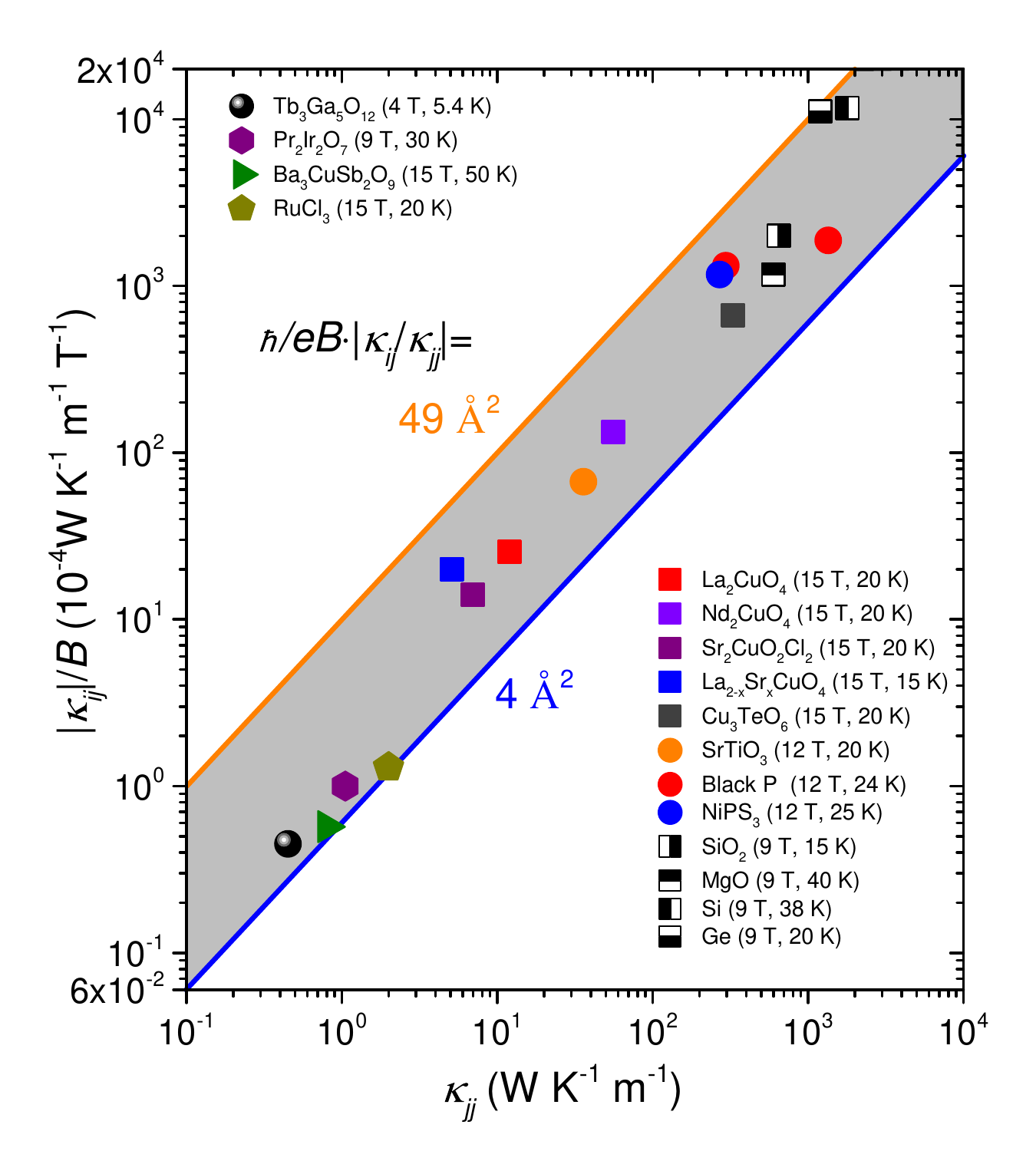}
		\caption{Thermal Hall angle in different insulators. The transverse thermal conductivity divided by magnetic field as a function of longitudinal thermal conductivity in different insulators (source:~\cite{Strohm2005,Sugii2017,Li2020, Grissonnanche2020,Boulanger2020,Akazawa2020,Chen2022,Uehara2022,Lefran2022,Li2023,Meng2024,Jin2024}). Note that magnetic insulators with a non-linear field dependence of the thermal Hall effect, such as Lu$_2$V$_2$O$_7$\cite{Onose2010} and Cu(1-3, bdc)\cite{Hirschberger2015}, are not included in the figure.}
		\label{fig:comparision}
	\end{figure}

	Experimentally, the thermal Hall angle  at $B$ = 10 T peaks to $\approx  10^{-2}$ in black P. The length scale extracted from this angle and fundamental constants ($\lambda_{\mathrm{tha}}$ = $\ell_B \sqrt{\kappa_{ij}/\kappa_{jj}}$) is about 5 \AA. Intriguingly, in all insulators displaying a thermal Hall signal, this length scale lies in a narrow range  ~\cite{Li2023,Meng2024} (2 \AA $~<\lambda_{\mathrm{tha}} <7$ \AA). As seen in Fig.~\ref{fig:comparision}, this remains true for recently studied cases of Si, Ge and quartz \cite{Jin2024}.
    
	Suppose an electric dipole moment of $\delta p$ traveling with the velocity of sound, $v_\mathrm{s}$ (loosely linked to the Debye frequency by $v_\mathrm{s} \sim a \omega_\mathrm{D} $, where $a$ is the interatomic distance).  The exerted torque, that is the angular derivative of the energy, would be $B  a \omega_D \delta p$.  Divided by the Debye energy, it yields $\theta_H\simeq\frac{a \delta p}{\hbar} B$, the order of magnitude of the expected rotation angle. With $\ell_B^2=\frac{\hbar}{e B}$,  assuming  $\delta p \approx e a$, leads to:
	
	\begin{equation}
		\theta_H\approx\frac{a^2}{\ell_B^2}
		\label{angle-2}
	\end{equation}
	
	This is of the order of magnitude of the measured signal. A rigorous treatment would presumably include the Gr\"uneisen parameter \cite{Ashcroft76} and the atomic electric dipole polarizability. The former, which is dimensionless, remains of the order of unity and the latter introduces a length scale which does not vary widely among different solids \cite{Schwerdtfeger2019}. This admittedly hand-waving scenario, attributing the thermal Hall signal to the interaction between a static magnetic field and traveling electric dipolar waves, was not considered before. Moreover, its influence merits in-depth consideration when applied to other elemental insulators.
	
	\section{Conflict of interest}
	The authors declare that they have no conflict of interest.

	\section{Acknowledgments}
	We thank Beno\^it Fauqu\'e for stimulating discussions. This work was supported by the National Key Research and Development Program of China (2022YFA1403500), the National Natural Science Foundation of China (12004123, 51861135104 and 11574097 ), and the Fundamental Research Funds for the Central Universities (2019kfyXMBZ071).  X. L. was supported by The National Key Research and Development Program of China (2023YFA1609600) and the National Natural Science Foundation of China (12304065).
	
	 \section{Author contributions}
	Xiaokang Li, Zengwei Zhu,  and Kamran Behnia proposed and supervised the project. Xiaokang Li and Xiaodong Guo prepared the samples, designed and performed the measurements.  Xiaokang Li, Zengwei Zhu,  and Kamran Behnia analyzed data and wrote the manuscript with the contributions of all authors.
	
	\bibliography{main}
	\clearpage

\begin{center}{\large\bf Supplementary Materials for ``Angle-dependent planar thermal Hall effect by quasi-ballistic phonons in black phosphorus" by X. Li et al.}\\
\end{center}

\renewcommand{\thesection}{S\arabic{section}}
\renewcommand{\thetable}{S\arabic{table}}
\renewcommand{\thefigure}{S\arabic{figure}}
\renewcommand{\theequation}{S\arabic{equation}}

\setcounter{section}{0}
\setcounter{figure}{0}
\setcounter{table}{0}
\setcounter{equation}{0}

\section{Materials and Methods}
Black phosphorus crystals used in this work are same to our previous report~\cite{Li2023}. They were synthesized under high pressure and came from two different sources. Sample \#1 was obtained commercially and sample \#2 was kindly provided by Prof. Yuichi Akahama (University of Hyogo). 

All thermal transport experiments were performed in a commercial measurement system (Quantum Design PPMS) within a stable high-vacuum sample chamber. An one-heater-four-thermocouples (type E) techniques was employed to simultaneously measure the longitudinal and transverse thermal gradient. The thermal gradient in the sample was produced through a 4.7 k$\Omega$ chip resistor alimented by a current source (Keithley6221). The DC voltage on the heater and thermocouples (thermometers) was measured through the DC-nanovoltmeter (Keithley2182A). The thermocouples, the heat-sink, and the heater were connected to samples directly or by gold wires with a 50 microns diameter. All contacts on the sample were made using silver paste. In the angular dependent measurements, a thermal transport rotation probe was used.

The longitudinal ($\nabla_i T$) and the transverse ($\nabla_j T$) thermal gradient generated by a longitudinal thermal current $J_q$ were measured. They lead to the longitudinal ($\kappa_{ii}$) and the transverse ($\kappa_{ij}$) thermal conductivity, as well as the thermal Hall angle ($\nabla_j T / \nabla_i T$):
\begin{equation}\label{kappaii}
\kappa_{ii} = \frac{Q_i}{\nabla_i T}
\end{equation}
\begin{equation}\label{thermal-Hall-angle}
\frac{\nabla_j T}{\nabla_i T} = \frac{\kappa_{ij}}{\kappa_{jj}}
\end{equation}
\begin{equation}\label{kappaij}
\kappa_{ij} = \frac{\nabla_j T}{\nabla_i T} \cdot \kappa_{jj}
\end{equation}
Here $Q$ is the heat power.

\section{Raw data of angular dependent planar thermal Hall effect}  
\begin{figure*}[ht]
\includegraphics[width=1\linewidth]{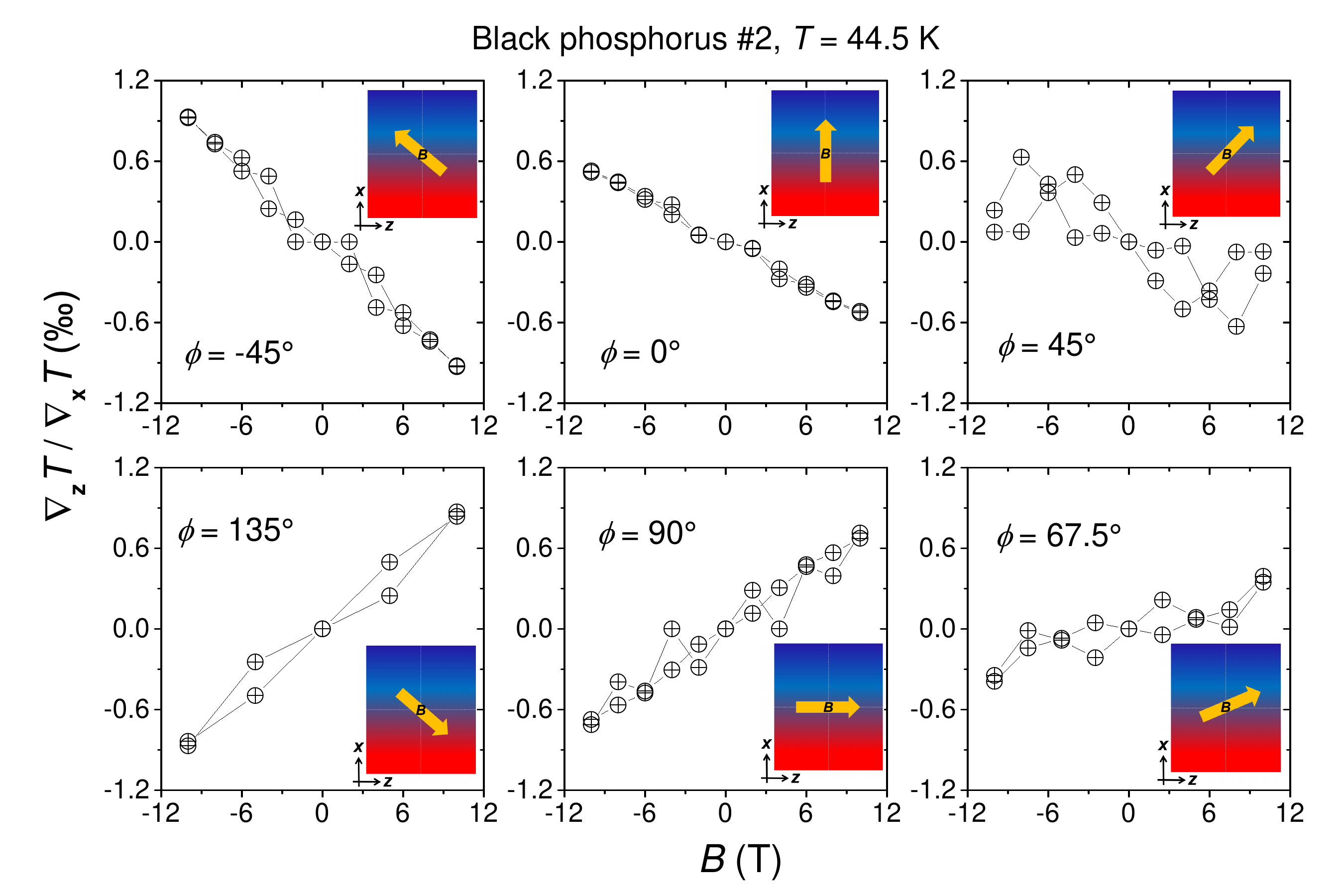} 
\caption{\textbf{Raw data of angular dependent planar thermal Hall effect in black phosphorus.} Field dependent thermal Hall angle ($\nabla_z T/\nabla_x T$) at a constant temperature and six different field orientations.}
\label{fig:raw-data-PTHE}
\end{figure*}

Fig.~\ref{fig:raw-data-PTHE} shows the field dependent thermal Hall angles ($\nabla_j T / \nabla_i T$) with  the $J_q$ alongs $x$ axis and the $(\nabla T)_{\perp}$ along $z$ axis at six different field orientations. $\phi$ is the angle of the magnetic field with respect to the $J_q$ ($x$ axis). $\phi = 0^\circ$ and $\phi = 90^\circ$ represent the orientations along two high-symmetry axes, and their $\nabla_j T / \nabla_i T$ signal have the same amplitude but opposite signs. $\phi = -45^\circ$ and $\phi = 45^\circ$ represent two diagonal orientations of the puckered honeycomb plane that have the maximum and vanished responses respectively, suggesting that the planar thermal Hall signal consists of two distinct contributions each along one high-symmetry axis. $\phi = -45^\circ$  and $\phi = 135^\circ$ have signals with opposite signs, as expected from the antisymmetric operation ($\kappa_{ij}$(B) = -$\kappa_{ij}$(-B)).


\section{Torque exerted by the magnetic field on moving dipoles}
The torque exerted by a static magnetic field, $\overrightarrow{B}$, on an electric dipole of $\overrightarrow{P}$ moving with a velocity of $\overrightarrow{v}$ is :
\begin{equation}
\begin{split}
\overrightarrow{\tau} &= \overrightarrow{P}\times(\overrightarrow{B}\times\overrightarrow{v}) \\
                      &= (\overrightarrow{v}\times\overrightarrow{B})\times\overrightarrow{P} \\
                      &= (\overrightarrow{P}\cdot\overrightarrow{v})\overrightarrow{B}-(\overrightarrow{P}\cdot\overrightarrow{B})\overrightarrow{v}
\end{split}
\label{S4}
\end{equation}

The electric dipole $\overrightarrow{P}$, field $\overrightarrow{B}$, and the velocity $\overrightarrow{v}$ can be expanded as:
\begin{equation}
\begin{split}
\overrightarrow{P}= P_x\overrightarrow{e_x}+P_y\overrightarrow{e_y}+P_z\overrightarrow{e_z};\\ \\
\overrightarrow{B}= B_x\overrightarrow{e_x}+B_y\overrightarrow{e_y}+B_z\overrightarrow{e_z};\\ \\
\overrightarrow{v}= v_x\overrightarrow{e_x}+v_y\overrightarrow{e_y}+v_z\overrightarrow{e_z}.
\end{split}
\label{S5}
\end{equation}

Thus, the \ref{S4} can be rewritten as: 
\begin{equation}
\begin{split}
\overrightarrow{\tau}= (P_x v_x+P_y v_y+P_z v_z)\cdot (B_x\overrightarrow{e_x}+B_y\overrightarrow{e_y}+B_z\overrightarrow{e_z})\\ \\
-(P_x B_x+P_y B_y+P_z B_z)\cdot (v_x\overrightarrow{e_x}+v_y\overrightarrow{e_y}+v_z\overrightarrow{e_z})\\ \\
= (P_y v_x B_x + P_z v_z B_x-P_y B_y v_x-P_z B_z v_x)\cdot \overrightarrow{e_x} \\ \\
+ (P_x v_y B_y + P_z v_z B_y-P_x B_x v_y-P_z B_z v_y)\cdot \overrightarrow{e_y}  \\ \\
+ (P_x v_x B_z + P_y v_y B_z-P_x B_x v_z-P_y B_y v_z)\cdot \overrightarrow{e_z}
\end{split}
\label{S6}    
\end{equation}

\section{The angular dependence of the torque and the observed bi-axial symmetry}

In our experiment, the magnetic field was rotated in the $xz$ plane, the longitudinal thermal gradient was applied along the $x$-axis. Putting $B_y=0$ and $P_y=0$, Equation \ref{S6} becomes:

\begin{equation}
\begin{split}
\overrightarrow{\tau}= (P_z v_z B_x-P_z B_z v_x)\cdot \overrightarrow{e_x} \\ \\
- (P_x B_x v_y + P_z B_z v_y)\cdot \overrightarrow{e_y}\\ \\
+ (P_x v_x B_z - P_x B_x v_z)\cdot \overrightarrow{e_z}
\end{split}
\label{S7}    
\end{equation}

We measured the transverse thermal gradient along the $z$-axis. The torque along the z-axis is equal to :
\begin{equation}
\tau_z= (P_x v_x) B_z - (P_x v_z)B_x
\label{S8}    
\end{equation}

One can see that the expected torque along the $z$-axis has two components. Each of them vanishes when the field is along either the $x$-axis or the $z$-axis. This is in agreement with the experimental data seen in Figure 3 of the main text.

\begin{figure*}[ht]
\includegraphics[width=1\linewidth]{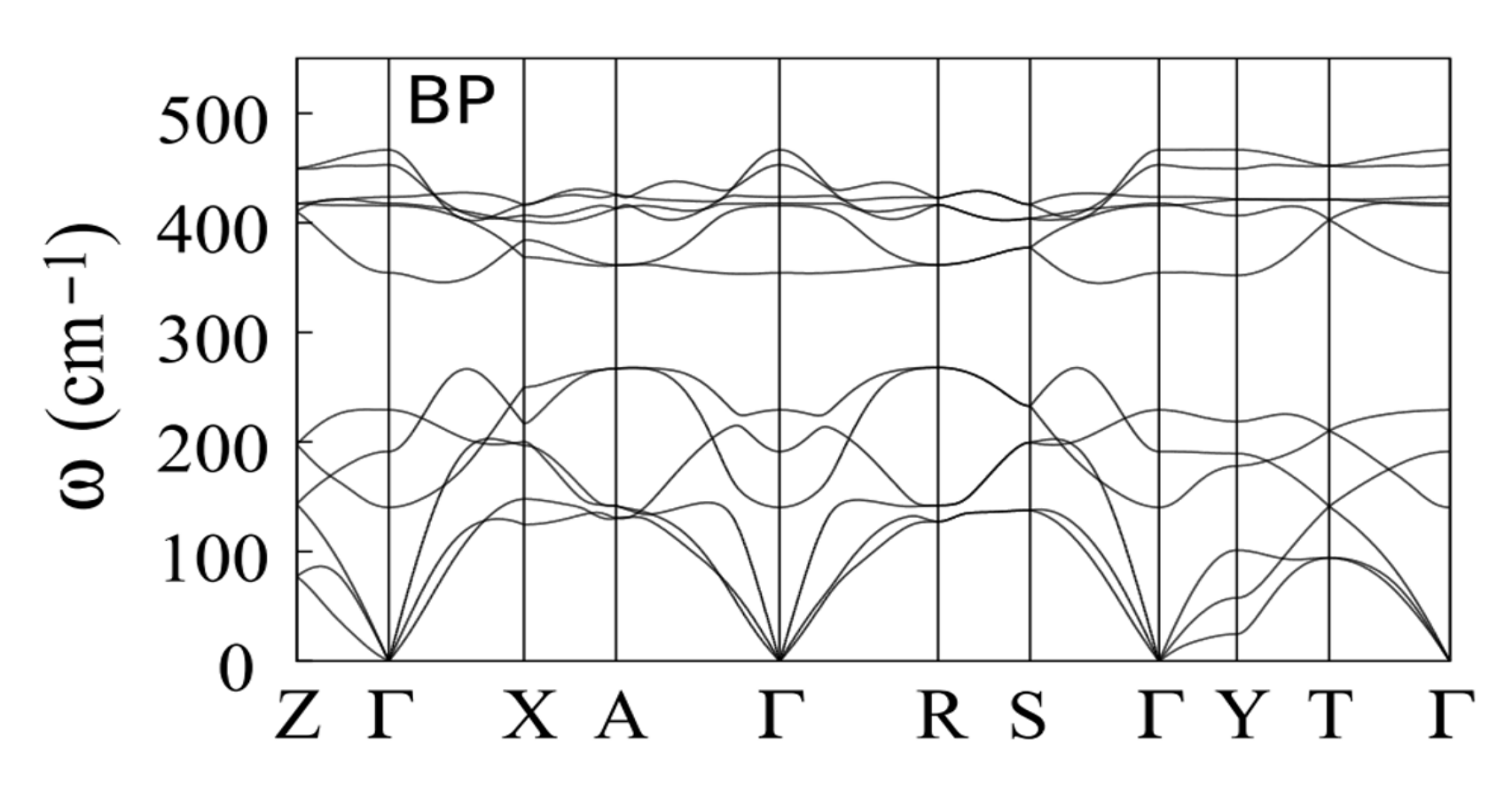} 
\caption{\textbf{Phonon spectrum in black phosphorus.} Adapted from \cite{Machida2018}.}
\label{fig:phonon-spectrum}
\end{figure*}

\begin{table}[htb]
\centering
\begin{tabular}{|c|c|c|c|}
\hline
 & \textbf{Mode} & \textbf{Calc. Machida et al.} & \textbf{Exp. Fujii et al.} \\ 
\hline
\multirow{4}{*}{$v_x$} 
 & LA$_x$ & 0.454 & 0.46  \\  
 & TA$_z$ & 0.490 & 0.46  \\  
 & TA$_y$ & 0.119 & --  \\  
\cline{2-4}
 & Average & 0.354 & --  \\  
\hline
\multirow{4}{*}{$v_z$} 
 & LA$_z$ & 0.833 & 0.96  \\  
 & TA$_x$ & 0.488 & --  \\  
 & TA$_y$ & 0.287 & --  \\  
\cline{2-4}
 & Average & 0.536 & --  \\  
\hline
\multirow{4}{*}{$v_y$} 
 & LA$_y$ & 0.505 & 0.510  \\  
 & TA$_x$ & 0.160 & 0.170  \\  
 & TA$_z$ & 0.289 & 0.310  \\  
\cline{2-4}
 & Average & 0.318 & 0.330 \\  
\hline
\end{tabular}
\caption{\textbf{Phonon velocities in black P.} Velocities of the three acoustic phonon modes (in units of 10$^4$ m/s) along different high symmetry axes}
\label{table-phononvelocity}
\end{table}

\section{Phonon velocities and phonon spectrum in black Phosphorus.}
Table~\ref{table-phononvelocity} lists the velocity of longitudinal (LA) and transverse acoustic (TA) modes in black P according to the calculated phonon spectrum \cite{Machida2018} and the neutron scattering experiments \cite{FUJII1982579}. Note the good agreement between theory and experiment. Remarkably, the longitudinal mode is faster than the transverse mode.

Fig.~\ref{fig:phonon-spectrum} shows the phonon spectrum up to 500 cm$^{-1}$. The lowest optical phonon frequency of black phosphorus is around 140 cm$^{-1}$, much larger than the acoustic phonons below 40 cm$^{-1}$ which are the focus of this work.

\end{document}